# Sustaining Educational Reforms in Introductory Physics

Steven J. Pollock and Noah D. Finkelstein

Department of Physics
University of Colorado at Boulder
Boulder, CO 80309

While it is well known which curricular practices can improve student performance on measures of conceptual understanding, the sustaining of these practices and the role of faculty members in implementing these practices are less well understood.  We present a study of the hand-off of *Tutorials in Introductory Physics* from initial adopters to other instructors at the University of Colorado, including traditional faculty not involved in physics education research.  The study examines the impact of implementation of Tutorials on student conceptual learning across eight first-semester, and seven second-semester courses, for fifteen faculty over twelve semesters, and includes roughly 4000 students. It is possible to demonstrate consistently high, and statistically indistinguishable, student learning gains for different faculty members; however, such results are not the norm, and appear to rely on a variety of factors. Student performance varies by faculty background - faculty involved in, or informed by physics education research, consistently post higher student learning gains than less-informed faculty. Student performance in these courses also varies by curricula used –all semesters in which the research-based *Tutorials* and Learning Assistants are used have higher student learning gains than those semesters that rely on non-research based materials and do not employ Learning Assistants.

# Introduction:

It is increasingly well understood how to promote student learning in large-enrollment, introductory physics courses. At least by measures of performance on conceptual surveys, a wide variety of pedagogical approaches based on the findings of physics education research have been demonstrated to dramatically improve student performance [1,2, 3]. Two of the great remaining educational challenges are how to scale reforms beyond the location where they were developed originally, and whether these reforms can be sustained. There is significant evidence that it is possible to replicate the successes of curriculum developers [4, 5,6]; however success is far from guaranteed [7,8]. It appears that how these practices are enacted is critical. Notably while interactive engagement (IE) courses outperform nearly all non-IE classes [9], the variation in normalized learning gains among IE courses is dramatic. This variation of success among implementations points to some of the difficulties of sustaining educational change [10,11]. What happens when proven strategies are handed off to faculty not involved in the design or initial implementation of educational reforms?

This paper explores the role of sustaining reforms that have been demonstrated to be successful both by the original creators and by the initial implementers at a second institution. We address the following research questions:

- Is it possible to hand off these educational reforms from one faculty member to another and sustain the level of student achievement realized initially?
- Does the instructor's background matter (physics education researcher (PER) *vs* non-PER)?
- Do the particular curricula matter, or is it entirely a matter of faculty practices?
- Can we demonstrate the development of faculty members as they implement these new pedagogical practices?

## Background, Environment, and Data Sources:

In prior work [6] we described our implementation of *Tutorials in Introductory Physics* [12] and demonstrated that it was possible to replicate measurements of conceptual shifts in students that were remarkably similar to those of the curricular authors, researchers at University of Washington (UW) Physics Education Group. Figure 1 illustrates this similarity. Data from UW are shown for post-instruction results without the use of Tutorials (blue), and following their introduction (red) [13]. Results from first implementation of Tutorials at the University of Colorado at Boulder (CU) are shown in yellow. Each category represents conceptual questions from the UW [13] used to demonstrate the effectiveness of their Tutorial materials.

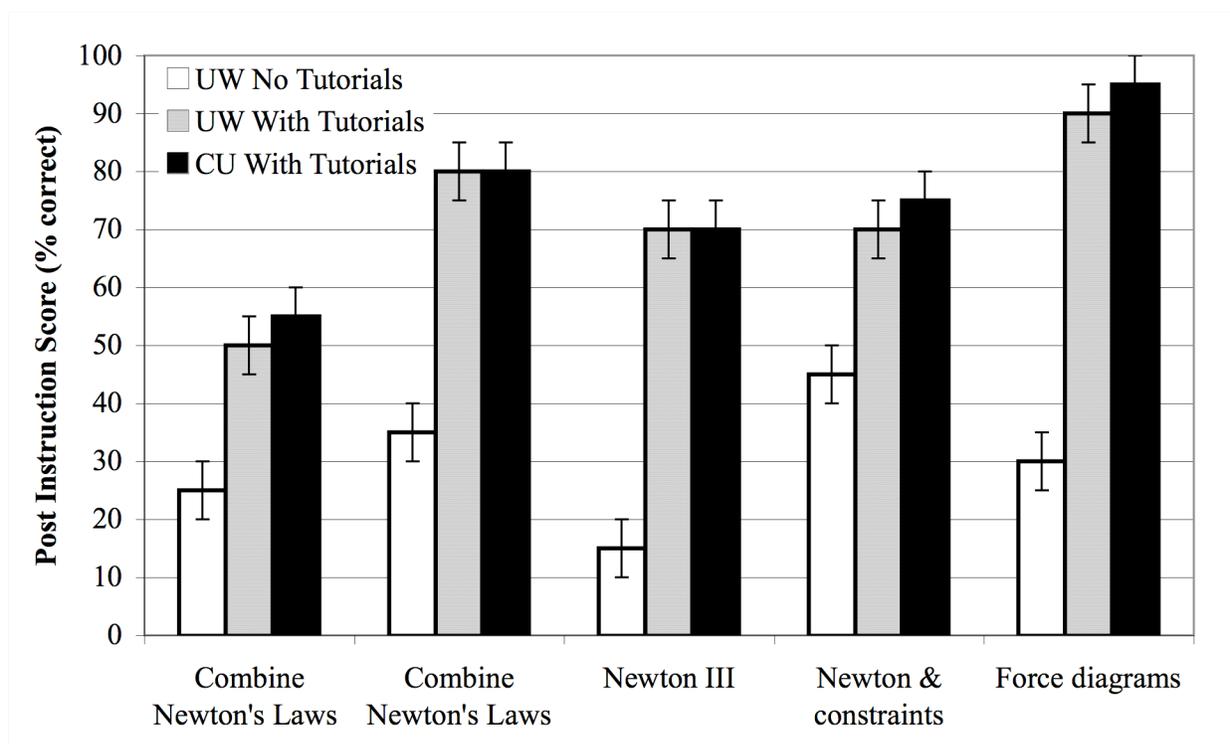

**Figure 1:** Post-test results on four published [13] conceptual questions asked on midterm exams. Results from UW and CU are shown. "No-Tutorial" scores are shown for courses at UW only. Following UW reporting practices our results are rounded to the nearest 5%.

To understand how we achieved such similar results, our prior analysis [6] examined various nested levels, or frames of context [14], ranging from the specific tasks that students engaged in, to the surrounding departmental and institutional structures. Two critical frames were the *situation* and the *class culture*. "Situations" refer to the surrounding structure of students working at Tutorials, including the rooms, teaching assistants and materials and social setting during the 50-minute recitation sessions. The "class culture" is defined as the environment in which these situations are set- the norms of behavior for the course, the purposes of the Tutorials, grading policies, etc. While the specific tasks, the Tutorial workbook problems, may be straightforward to hand off from one faculty member to another, the situation and class culture levels are harder to prescribe, and thus potentially subject to greater variation. As the leading role for these courses is handed from one faculty member to another, so too might we anticipate shifts in the enacted practices that constitute situations and class culture. Here, we begin to explore these issues in studies of the sustained implementation of Tutorials and related course reforms, including *Peer Instruction*[15] in our introductory course sequence.

*Course Description:*

The studies at the University of Colorado (CU) occurred in the calculus based, large-scale (N=300 to 600) introductory physics sequence. The student population is a mix of engineering majors (50%), natural science majors (20%) and a variety of others, including undeclared majors. The course is 75% male, and over half the students are freshmen. Data were collected in ten first-semester courses (Fall 2001, then Fall 2003 - Fall 2007) and seven second-semester courses (Fall 2004 - Fall 2007). All instructors in this study since Fall 2003 used Concept Tests [15] during three 50-minute lectures per week in a 200- to 300-student classroom, online homework systems [16,17], one 50-minute per week recitation section, and a staffed help-room, which was voluntary for students. The introductory laboratory experience is decoupled from this course, and typically taken concurrently with the second semester lecture course. All second-semester courses, and many of the first-semester courses, implemented Washington's Tutorials [12] during the weekly recitation sections. The remaining first-semester courses ran more traditional recitation sections, as discussed below. The recitation sections were staffed by one departmentally assigned teaching assistant, TA, and in the case of Tutorials, an additional undergraduate assistant, a Learning Assistant (discussed below), was added. Two faculty members are generally assigned to teach these large courses; one is in charge of lectures, the other in charge of Tutorials and the rest of the course administration. The faculty member lecturing is historically the lead of the two-person team, setting course pace, goals and overall workload. The "secondary" faculty member is typically in charge of the course management, TAs, logistics and the recitation sections, where Tutorials were implemented. As such, the lead faculty is dominant in establishing the "class culture" and the secondary instructor establishes and frames the "situations" for Tutorials. CU physics faculty are cycled through these classes; during the course of this study, eleven different faculty members participated in the first semester course and eight faculty participated in the second semester course. Faculty in this study are categorized by their familiarity with physics education research, the underpinning principles of learning theory and practices built from them. PER faculty members are defined as members of the PER-group, have conducted research in student learning, published and taught classes in the field. "Informed" faculty are those who are familiar with the field of physics education research, some of its findings, and have engaged in workshops led by PER faculty, participated in departmental brownbag discussions, and/ or engaged in readings from the field. These faculty are observed to enact some of the principles from the field, such as those described by Redish [1]. Traditional faculty are the typical university faculty members who receive little or no preparation in teaching, may well report knowing about the field [10,11], have not participated in departmental activities designed to inform faculty about PER, and are not observably enacting the principles of the field. Faculty are denoted by "P" (PER researchers), "I" (informed) and "T" (traditional), respectively[18]. The initial implementations of Tutorials in both the first and second semester courses were led by PER research faculty. Following, there were only two other cases (both in the second-semester course) where another PER-faculty member directed the course.

We attempted to implement a fairly high-fidelity replication of the University of Washington Tutorials [12], but local departmental and institutional constraints inevitably resulted in a wide variety of differences. Our recitation sections are all held on one day of the week. We created three temporary spaces (3-walled bays) in the basement laboratory areas, where we fit seven tables of 4 students each. Sections run in parallel, resulting in a somewhat noisy and cramped environment. Sections are staffed by a graduate Teaching Assistant (TA), and one or two undergraduate Learning Assistants (LAs). The LAs are students who had previously performed well in this class, expressed interest in teaching and became part of a broader program to support their development as teachers [19]. TAs and LAs are prepared in a weekly meeting several days prior to the Tutorial, where they look over student responses to online pretests to generate discussion of incoming student ideas, and are then guided through the upcoming Tutorial itself, working in small groups just as students do. When secondary faculty were not members of the PER group, the PER faculty worked fairly closely with these secondary instructors for the first week or two to try to explain the methodology and goals behind the Tutorials. We have created a written faculty guide, which provides weekly tips and suggestions, along with pretest responses from previous terms. After the first few weeks, faculty are left on their own to run these preparatory meetings as they see fit.

*Data collection:*

Each term, we administer pre- and post-instruction assessments of student conceptual understanding. The Force and Motion Concept Evaluation FMCE [20] was used in the mechanics course, generally issued the first and one of the last weeks of the term in recitation. Students are informed that the exam is diagnostic and does not impact their grades, but we encourage them to try their best. We obtain matched pre-post scores for roughly 60% of students enrolled in the class each term. We use the Brief Electricity and Magnetism Assessment BEMA [21] instrument for second semester physics, with typically 75% pre-post matched return rates. Notably, these instruments provide only a coarse evaluation of course achievements, and only represent one dimension of student achievement. Nonetheless, these evaluations allow us to compare, in aggregate, the impact of different course implementations by different faculty on the overall performance of students enrolled in these courses.

## Studies of Sustaining Tutorials

*Initial hand-off and intra-institutional replication*

The initial implementation of the Tutorials in each of the first and second semester calculus-based physics courses was conducted by the lead author [SJP], denoted P1. The Tutorials were implemented in the first semester sequence (mechanics) for the first time in fall 2003, and implemented for the first time in the second semester sequence (electricity and magnetism) in fall 2004. The following semester, spring 2005, another faculty member of the PER group, and nationally recognized educator taught the second semester course. This is the first instance of hand-off of the Tutorial curriculum at CU and arguably conducted under extremely favorable circumstances. The second instructor, P2, had assisted with the initial implementation of the Tutorials, is well versed in physics education research, and read many of the UW papers. He understood the underpinning pedagogy and purpose of the Tutorials. When P2 led the course, Sp05, an "informed" faculty member, I4, served in the secondary role. The course structures themselves were nearly identical from Fa04 to Sp05, but there was some variation in pacing, particular lecture notes, some concept tests, and text (Knight [22] vs. Halliday Resnick and Walker [23]). Tutorials were implemented in nearly identical manner, with teaching and learning assistants, weekly preparatory sessions and covering the same Tutorial topics.

Figure 2 shows distributions of pre and post-test score on the BEMA for both the initial implementation (dark) and the secondary instructor (light). The posttest score averages were the same, 59%, and the normalized learning gains indistinguishable 44% ± 1% (standard error on the mean) for the initial implementation and 43% ± 1%, secondary implementation.

*Standardizing curricular choice:*

While the lead author was implementing Tutorials in the second semester course, the first semester course was assigned to non-PER faculty. At that time, the Tutorials were new at CU, and not an institutionalized activity. Faculty members were not convinced of the differential effectiveness of Tutorials over traditional recitations, and there were no existing mechanisms to pay for extra costs of implementing this enhanced educational experience. As a result, faculty members were left on their own to select which activities would run in the recitation sections. As a matter of standard practice, the rest of the course activities, lectures with interactive concept tests, online homework system, and recitation sections of some sort, were continued and considered institutional practices at that point. The lead author (SJP) taught the first-semester course with Tutorials in Fall 03 and again in Spring 04. The following semester the new instructor opted to use the same pedagogical structure, small groups, and different curricular materials, selecting the workbook that came with the text [22]. A notable difference was that the institutional support for additional recitation leaders, in the form of Learning Assistants, did not exist. In recitation sections, the teacher to student ratio was roughly half what it was the prior semester, and the weekly preparatory sessions for recitation were more *ad hoc*, and certainly did not follow the UW approach [6,12]. The semester following, spring 05, the course instructor opted to revert back to more traditional recitation sections where TAs were often working at the board in front of the students reviewing homework problems or worksheets from the text or of their own design. Again, the recitations were led by single instructors rather than two individuals, and the weekly preparatory sessions were limited and less structured. Table 1 lists the lead/secondary roles, the type of recitation experience, number of students, post-test scores and the normalized learning gains for each of the first-semester courses studied. The two most significant variations among these semesters are the backgrounds of the instructors, and recitation experiences for students [24].

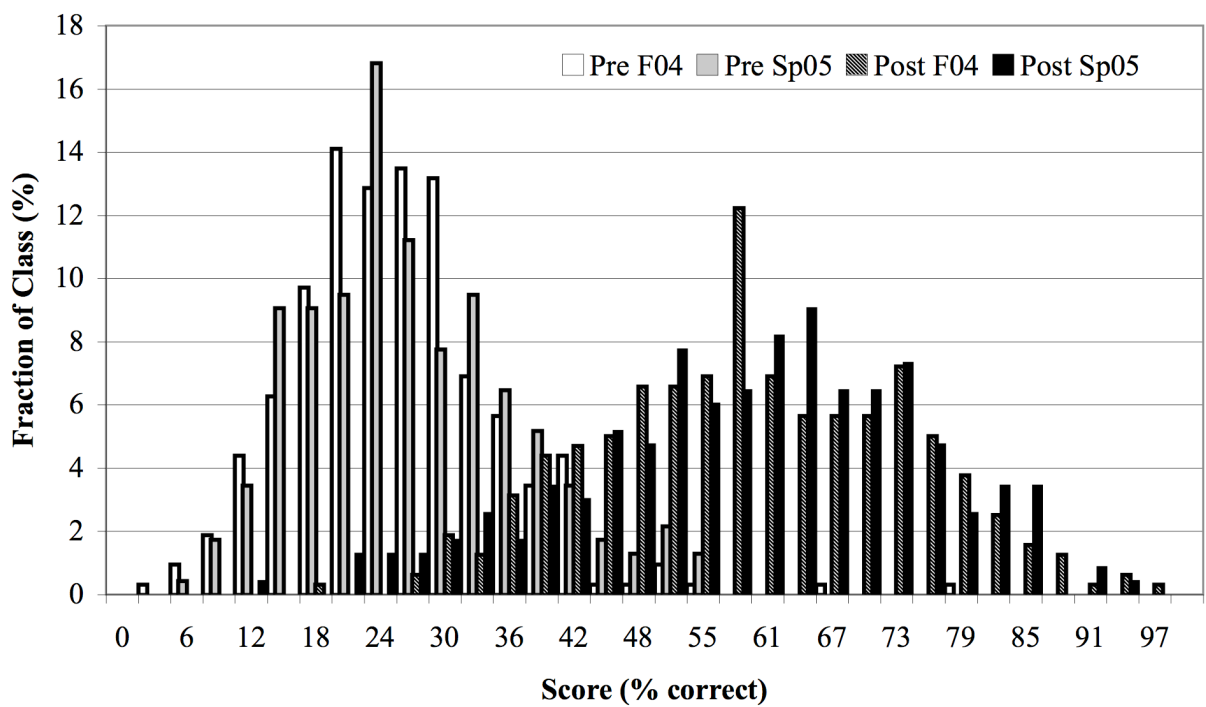

**Figure 2:** BEMA results pre (left) and post (right), shown as % correct. The data show two semesters of results in different shades. The term-to-term reproducibility of these histograms is striking. The two semesters were taught by different faculty, both PER researchers.

**Table 1: Summary of FMCE normalized gains over eight semesters of data collection.**

| Semester | Faculty Background Lead/Secondary | Recitation | N (matched) | Average posttest score * | Normalized gain <g> ** |
|---|---|---|---|---|---|
| F01 | T0 (alone)*** | Traditional | 265 | 52 | 0.25 |
| F03 | P1 (alone) | Tutorials | 400 | 81 (FCI data) | 0.63 |
| S04 | P1 (alone) | Tutorials | 335 | 74 | 0.64 |
| F04 | I1/T1 | Workbooks | 302 | 69 | 0.54 |
| S05 | T1/T2 | Traditional | 213 | 58 | 0.42 |
| F05 | T2/T3 | Traditional | 293 | 58 | 0.39 |
| S06 | I2/T4 | Tutorials | 278 | 60 | 0.45 |
| F06 | I3/I1 | Tutorials | 331 | 67 | 0.51 |
| S07 | I4/T4 | Tutorials | 363 | 62 | 0.46 |
| F07 | I2/I5 | Tutorials | 336 | 69 | 0.54 |

\* Average posttest score is a percent, scored using Thornton's recommended FMCE rubric [3], for those students with matched pre-post data. (Standard error of the mean is between 1-2% for all terms). Only F03 used the FCI[25] exam pre and post, all other terms are FMCE. (Note the similar gains for F03 and S04 on these two exams.)
\*\* Normalized gain in the last column is computed as the gain of the average pre and post scores for matched students. (Standard error of average gains is roughly ±.02 for all terms.)
\*\*\*The F01 course included traditional recitations, and peer instruction with colored cards in lecture.

Table 1 and Figure 3 summarize the overall measures of student conceptual learning gains in first semester courses. These allow for comparisons of Tutorial with traditional recitation sections. Figure 3 is a reproduced plot from Hake[9], (with permission) a histogram of student normalized learning gains for different environments. The red data represent traditional courses, the blue self-described "interactive engagement" environments. Overlaying this plot are arrows denoting the span of learning gains for Tutorial and Traditional recitation experiences offered at CU, as detailed in Table 1 [26, 27]. Notably all of the CU courses, whether traditional or Tutorial recitations, are interactive environments, using Peer Instruction and personalized, online homework systems. While CU courses span nearly the entire range of learning gains documented for IE courses elsewhere [9], we note that at CU, all courses with Tutorial experiences lead to learning gains higher than all classes that have traditional recitation experiences. Furthermore there is sizable variation of success among these implementations, suggesting a potential effect of faculty in these environments.

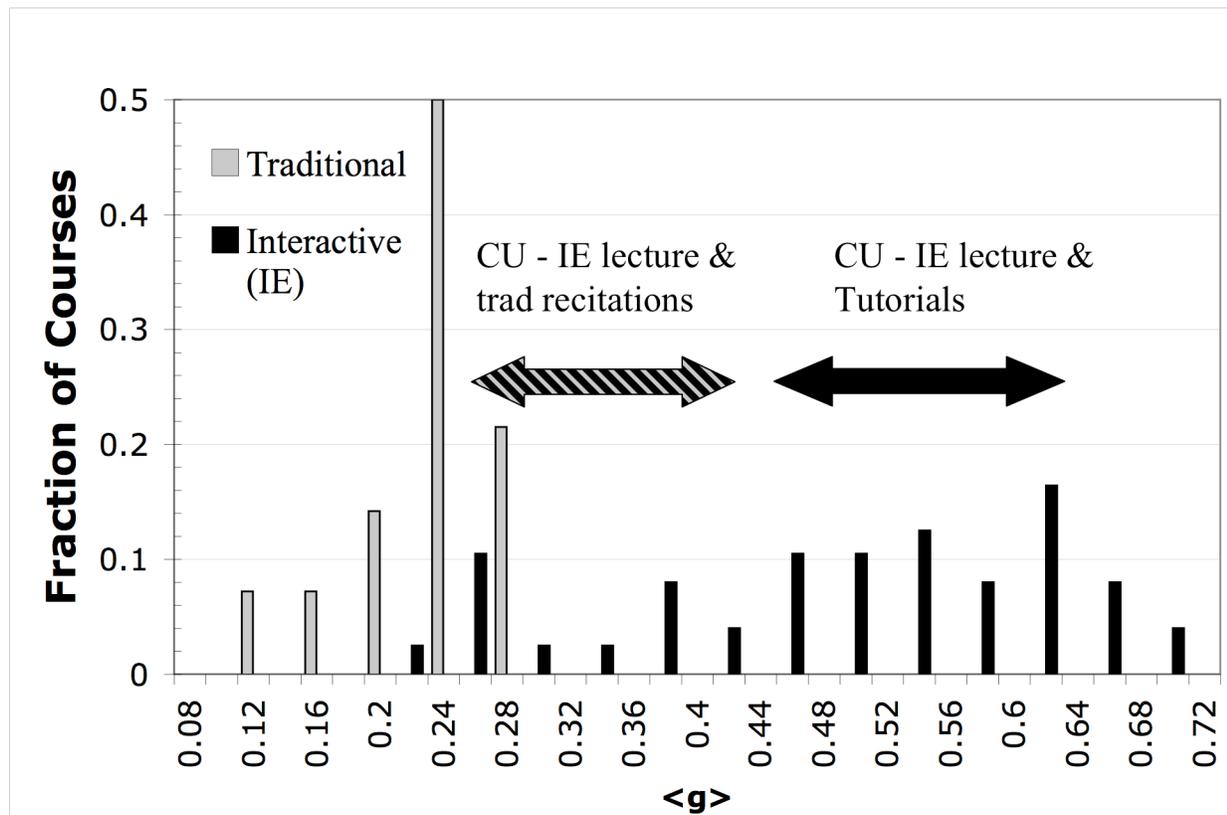

**Figure 3:** Plot of Normalized Learning Gains. Data from Hake[9] show FCI gains (background bars) Data for CU are FMCE gains, showing the span of CU interactive engagement (IE) lectures using Tutorials (right arrow), and IE lectures with *traditional* recitations (left arrow).

From spring 2006 to fall 2007 (and ongoing) in the first semester course, and from fall 2004 to fall 2007 (and ongoing) in the second semester course, Tutorials have been implemented consistently in the recitation section, including the increased teacher: student ratio by using LAs. While there are many conditions that have led to the institutionalization of Tutorials, from our multi-year history of involvement, we hypothesize critical features that have supported adoption of these materials may include:

- *Open discussions with faculty* about course goals and the impact of Tutorials;

- *Increased support for the recitation sections.* The number of Learning Assistants allocated to the department increased, initially through grant funding (APS, AIP, and AAPT PhysTEC funding [28] coupled to an NSF STEM-TP grant [29]). Subsequently the increased support came from the department and dean, and now university level support. Roughly $50k / yr supports the learning assistant program in the physics department.[30] Furthermore, the section size was decreased from 32 to 28 students, following requests from the faculty and department chair to the dean, based on the data presented above.

- *The use of consistent and validated measures of student learning.* By using the FMCE and BEMA surveys of conceptual mastery, collecting and presenting data such as those included in this paper, various stakeholders (faculty, department chair, dean and provost) were convinced that these programs were valuable for promoting student learning.

The detailed mechanism of why and how these programs are institutionally supported are left for another study.[31]

*Toward sustainability: intra-institutional replication to non-PER faculty*

After the second semester (Electricity and Magnetism) introductory course was successfully implemented and handed off to a second PER faculty member (spring 2005), it was handed off to a more traditional, but "informed" faculty member. This faculty member had been serving in the secondary role in the course during the spring 2005 semester, by running the weekly LA/TA Tutorial preparation sessions. In the fall 2005 this instructor, I4, assumed lead role in the course. Instructor I4 was provided all the course materials used by the PER-instructor,

P2, who led the class the prior term. These materials included lecture notes, concept tests, exams, homework, and Tutorial preparation materials. Observations and interviews with this faculty member documented that he primarily used these materials without modification [31].

Figure 4 plots the pre and post scores of student performance on the BEMA in this implementation; the fall 05 results are the lighter colored lines. (Posttest results are shown with dashed lines) The bar plots (shown in the background) are the same data shown in figure 2, more coarsely binned for comparison. In fall 06 the same professor, I4, taught this course once again (dark lines). This time the professor had the assistance of a PER faculty member, P1, serving in the secondary role. The pre- to –post normalized learning gains are 0.33 ± .01 (in Fall 05) and 0.40± .01 (in Fall 06). These should be compared to the 0.44 ± 0.01 gains of the PER instructors previously.

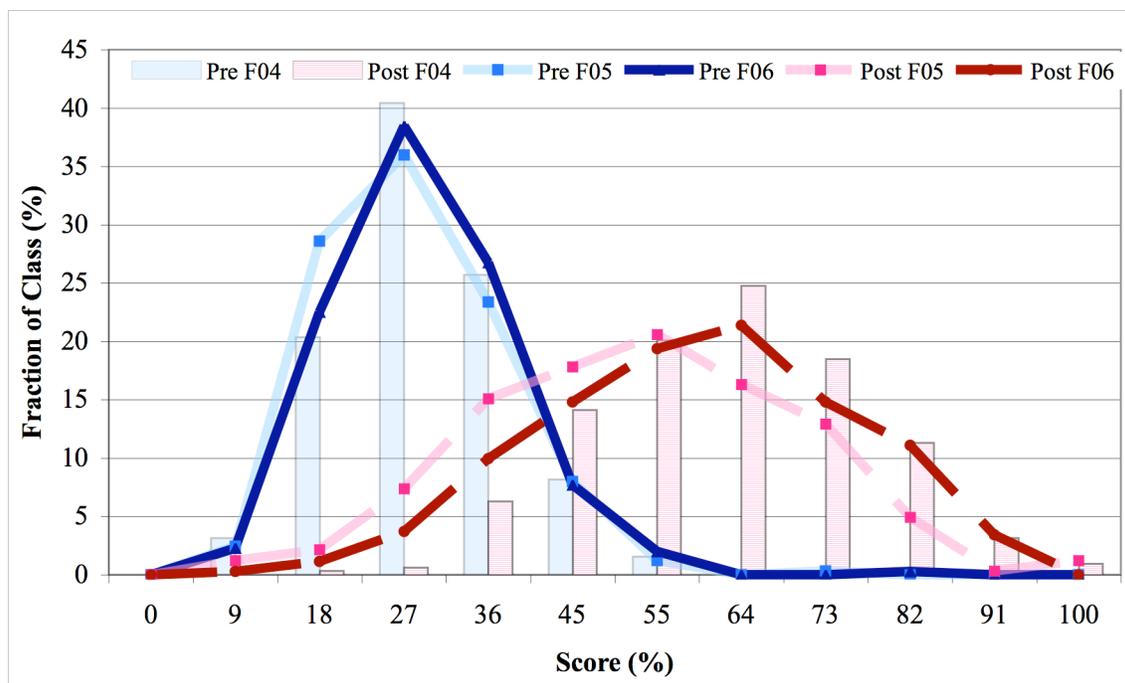

**Figure 4:** (Color) Pre (on the left) and Post (right) BEMA scores for second semester intro physics. Bars (shown lightly, in background) are coarsely binned data taken from figure 2, representing student performance in classes taught by PER faculty. Line plots are pre (solid) / post (dashed) data for Instructor I4 during two different implementations. Lighter lines are the first implementation, darker lines are his second time teaching this course.

Table 2 summarizes student performance in the second semester implementation of Tutorials, and the associated BEMA post-test scores and normalized learning gains. (Pre-test scores are not shown as they are extremely consistent from term to term, 26 ±1%) Data shown include results for hand-off from one PER faculty member to another (those data shown in Figure 2), and to "informed" and "traditional" faculty members. Notably there is a high level of fidelity where similar results are seen for the same faculty (Fa04 and Fa07) over a span of several years. As seen in Table I, the most significant variations among semesters are associated with the backgrounds of the instructors.

**Table 2: Summary of BEMA scores over seven semesters of data collection.**

| Semester | Faculty Background Lead/Secondary* | N (matched) | Average posttest score (%) | Normalized gain <g> |
|---|---|---|---|---|
| F04 | P1/P2 | 319 | 59 | 0.44 |
| S05 | P2/I4 | 232 | 59 | 0.43 |
| F05 | I4 (alone) | 325 | 50 | 0.33 |
| S06 | I6/T5 | 196 | 53 | 0.37 |
| F06 | I4/P1 | 351 | 56 | 0.40 |
| S07 | I7/I3 | 261 | 53 | 0.37*** |
| F07 | P1/T6 | 161** | 61 | 0.47 |

All semesters shown used Tutorials in recitation. Standard error of the mean for this table are very close to those in Table 1, 1-2% on posttest, and roughly ±.02 for normalized gains.
* Coding scheme is continued from Table 1, with common faculty given the same code, so e.g. in S05, the secondary faculty, labeled I4 here, is the same person as the lead faculty (I4) in Table 1, Spring 2007.
** In F07, half the class took the CSEM [32] test, with similar results.
*** No pretest was given in S07, but the average pretest has been 26±1% every other term, so we used this to estimate the normalized gain for that term.

**Discussion:**

The data from our studies span ten first semester, and seven second semester courses, roughly 4000 students and thirteen total semesters of implementing the Tutorials in recitation sessions. From these data we may begin to draw some conclusions about our abilities to sustain these effective educational practices, the role of instructors, curricular approaches, and some suggestions about faculty development.

We note first that it is possible to hand off these curricular reforms. The initial replication of the UW success was previously documented [6], and the subsequent hand-off of this course transformation to another faculty member (Figure 2) shows remarkable fidelity. The aggregate results of student performance are statistically indistinguishable for the Spring 05 and Fall 04 semesters.[33] Notably this is not a *de facto* result, and other faculty implementing these curricula do not necessary show similar student learning gains, as seen in Tables 1 and 2. Furthermore, these are aggregate results and we see significant variation within individual questions on the BEMA exam and on individual common exam questions for the two PER-instructor implementations. These variations reveal fine-grained differences in content emphasis and student variation in the two semesters. We note that while there was variation in individual faculty practice at a fine-grained level, the implementation of the Tutorials was remarkably similar at the level of the situation and the course culture levels. Each of the two instructors, and their respective students, emphasized the importance of sense making, reasoning and discussion in Tutorial sections. The tasks (specific choice of which Tutorial chapter to cover) and the situations were the same - students working in groups at the tables knew what the activity was about. The course culture emphasized making sense of given content and the role of conceptual understanding, what researchers at Maryland refer to as framing [34]. Which specific messages are sent and how they are sent is subtle and left for future detailed analysis [31,35].

While it is possible to hand off curricular reforms with success, it is not guaranteed. There is some suggestion that this has to do with the background of the faculty member, and whether or not these individuals are well versed in teaching practice that is informed by physics education research. In order to implement a course practice, one must attend to specific details of the local environment, institutional specifics that tend to vary with time, and make a myriad of decisions that are not prescribed or even documented – instructors must adapt their approaches and associated curricular practices [11]. For faculty members well versed in the field of physics education research, and familiar with the development of the specific innovations, in this case Tutorials, this adaptation can happen in a manner that is informed and aligned with the curricular goals. These faculty members can maintain a course culture and create situations supportive of the established pedagogical goals. Here and elsewhere [36], we observe that PER innovations maintain their fidelity when handed off to PER faculty. Notably from the data, Tables 1 and 2, we observe that instances when PER faculty are involved in instruction, in either the lead or secondary role, students post the highest learning gains. When PER-informed faculty, "I"s, are involved in instruction, students post higher learning gains than when only traditional faculty are involved. An example is seen in Table 1 by comparing spring 06 and fall 07 implementations. The same lead instructor, I2, partnering with a traditional (sp06) or informed (fa07) secondary instructor yields significantly different results[37]. In all cases we observe the positive impact of the curricula and the interactive nature of the courses, but not the same degree of effectiveness. While students are engaged in the same tasks of working through the Tutorials, the meaning of these tasks may be different – the course culture and situations that students engage in shift, in subtle but substantial ways, resulting in differential performance on these conceptual surveys.

At the same time, there is significant indication that the specific curricular practices matter. Of course the most compelling study would be to control for faculty member and vary the curricula. However, we have not had opportunity to observe the same faculty member teaching in both a traditional and Tutorial course[38].Nonetheless, we do observe that all instances of courses that include traditional recitation sections lead to student normalized learning gains that are lower than any course that implements Tutorials, seen in Figure 3. One may argue that these results may arise in part from having a greater number of instructors in the room, or that the training and preparatory sessions for the recitation leaders is different. We agree. That is, we believe that part of these innovative curricular practices is an increased teacher: student ratio and structured preparation sections. We also believe the materials themselves, which embody both content, emphasizing concepts and student reasoning, and social practice, group work and explicit justification, are important. There is a possible counterpoint, that the semester (Fa04, first semester course) using Knight's workbook materials post significant learning gains, comparable to those seen with Washington Tutorials. We find this result to be intriguing and suspect it to be due to two main factors. First it is worthy of note that the social practices (of students working in groups justifying reasoning) and the content of the workbooks appear to be remarkably similar to the Tutorials. In these environments, students were encouraged and observed to work in groups[39]. Ultimately the situations that students engaged in that semester (Fa04) were similar to those of the other semesters that implemented Tutorials. Second, the course instructor, while not a member of the PER faculty, had taken it upon himself to attempt an implementation of SCALE-UP [5]. In addition to reading about those particular curricular innovations, he visited North Carolina State University, and learned about the structure, practice, and philosophy of the reforms. While he was ultimately not successful because of institutional constraints (not finding a suitable room), we believe this faculty member holds a greater understanding of the underpinning philosophy and approach to the Tutorials as a result of this attempt. We suspect the course culture that he created was in greater alignment with PER goals than any of the other non-PER faculty.

© 2008 Pollock and Finkelstein                    16                    to appear in *Physical Review: Special Topics PER*

Finally, there is some evidence that engaging the implementation of these reforms can promote learning – for the faculty. While not certain, the data in Figure 4 suggests that Professor I4 developed expertise at teaching, at least as measured by shifts in student performance on conceptual surveys. It may be the case that I4's beliefs about his role, the goals for the students and ultimately the course culture shifted as a result of teaching in his first attempt. These are corroborated by interviews with I4 over the course of his teaching in these two implementations [31]. A caveat to this interpretation is that I4 had additional resources in the form of a PER-versed secondary instructor, P1, in the second implementation – perhaps the greater instructional resources the second term led to greater learning for the students. Nonetheless, we suspect that this approach of modeling practices may well be effective and sustainable. An instructor new to these practices begins by serving in the secondary role with a well-versed faculty member leading the course (e.g. I4 in spring 05), then assumes a lead role for the course (fall 05) and refines his or her approach to teaching in these reformed manners (fall 06). Such an approach is a modification of the co-teaching approach, advocated and studied by Henderson [40].

**Conclusions:**

While it is well known which curricular practices can improve student performance on measures of conceptual understanding, the sustaining of these practices and the role of faculty members in implementing these practices are less well understood. We have examined the implementation of *Tutorials in Introductory Physics* at the University of Colorado over a four-year span, covering twelve different implementations of Tutorials by fifteen faculty. We observe that it is possible to demonstrate strong and consistent learning gains for different faculty members; however, such results are not the norm, and rely on a variety of other factors.

We observe that both faculty background and the particular curricula used in recitation sections are significant contributors to the variation in student performance from semester to semester.

**Acknowledgements:**

This work was conducted with the support of the National Science Foundation (CCLI, STEM-TP, CAREER, and CIRTL), and the AAPT/ AIP/ APS PhysTEC. Any opinions, findings and conclusions or recommendations expressed in this material are those of the authors and do not necessarily reflect the views of the National Science Foundation (NSF). We are particularly grateful to UW PEG, Valerie Otero, Chandra Turpen, the CU Faculty participating and opting to implement Tutorials, the Physics Education Research & Discipline-based Education Research groups at CU.

However, the *manner* in which these practices were implemented may in fact have varied considerably. An ongoing study examines the question of how faculty implement practices, but such an examination occurs at a finer-grain size than the present study, which focuses on curricular choices, and faculty background at a broader scale.